\documentclass[
reprint,
amsmath,
amssymb,
aps,
pra
]{revtex4-2}

\usepackage{graphicx}

\usepackage{dcolumn}
\usepackage{bm}
\usepackage{hyperref}

\usepackage{mathtools}
\usepackage{cleveref}

\usepackage{physics}
\usepackage{qcircuit}
\usepackage{tikz}
\usetikzlibrary{shapes,arrows,calc,positioning, automata,decorations.pathmorphing}

\DeclareMathOperator{\N}{\mathbb{N}}

\DeclareMathOperator{\C}{\mathbb{C}}

\DeclareMathOperator*{\argmin}{\arg\min}

\DeclareMathOperator*{\dg}{\dagger}

\newcommand{\A}{\mathcal{A}}
\newcommand{\convA}{\mathrm{conv}(\mathcal{A})}
\newcommand{\atom}[1]{\left\lVert#1\right\rVert_{\mathcal{A}}}

\newcommand{\Gu}{G_{\mu}} 

\newcommand{\Ham}{\mathcal{H}}

\begin{document}

\title{Super-resolution of Green's functions  on noisy quantum computers}

\author{Diogo Cruz}
    \affiliation{Instituto de Telecomunica\c{c}\~{o}es, Portugal}
    \affiliation{Instituto Superior T\'{e}cnico, Universidade de Lisboa, Portugal}
\author{Duarte Magano}
    \affiliation{Instituto de Telecomunica\c{c}\~{o}es, Portugal}
    \affiliation{Instituto Superior T\'{e}cnico, Universidade de Lisboa, Portugal}

\date{\today}

\begin{abstract}
    Quantum computers, using efficient Hamiltonian evolution routines, have the potential to simulate Green's functions of classically-intractable quantum systems.
    However, the decoherence errors of near-term quantum processors prohibit large evolution times, posing limits to the spectrum resolution.
    In this work, we show that Atomic Norm Minimization, a well-known super-resolution technique, can significantly reduce the minimum circuit depth for accurate spectrum reconstruction.
    We demonstrate this technique by recovering the spectral function of an impurity model from measurements of its Green's function on an IBM quantum computer.
    The reconstruction error with the Atomic Norm Minimization is one order of magnitude smaller than with more standard signal processing methods.
    Super-resolution methods can facilitate the simulation of large and previously unexplored quantum systems, and may constitute a useful non-variational tool to establish a quantum advantage in a nearer future.
\end{abstract}

\maketitle

The simulation of quantum systems beyond the capabilities of classical computers is one of the oldest and most important promises of quantum computing.
Since the seminal ideas of Manin \cite{Manin80} and Feynman \cite{Feynman82}, quantum algorithms have been designed for Hamiltonian evolution \cite{Lloyd1996}, estimating eigenvalues \cite{AbramsLloyd}, preparing ground states \cite{Farhi2000, ITE}, computing transition probabilities \cite{Preskill}, and studying equilibrium properties \cite{Terhal2000}.
More recently, there has been a growing interest on quantum algorithms for calculating  Green's functions \cite{Wecker, Bauer, Kreula, Jaderberg, Keen, Endo, Rungger,  Zhu, Jamet, Baker}.
These dynamic pair correlation functions play a central role in quantum theory, with applications in chemistry \cite{martin_reining_ceperley_2016}, condensed matter \cite{ManyBodytextbook}, and  high-energy physics \cite{QFTtextbook}.

Unfortunately, the Green's functions of many-body strongly-correlated systems are notoriously hard to compute classically \cite{Schuch2009}.
Leveraging quantum computers' capacity to store and time-evolve large wave functions, Refs. \cite{Wecker, Bauer, Kreula, Jaderberg, Keen, Endo} propose measuring Green's functions in the real time domain.
Other approaches exploit the Lehmann's representation \cite{Endo, Rungger,  Zhu}  and the continued fraction representation \cite{Jamet, Baker} of Green's functions.
Other works \cite{Keen, Endo, Jamet, Rungger, Zhu} consider variational methods, which are better suited to the current era of noisy intermediate-scale quantum (NISQ) computers than Hamiltonian evolution or phase estimation-based algorithms.
However, it is difficult to characterize the computational scaling of such variational algorithms, since in general they may be offloading the hardness of the problem onto the optimization step \cite{McClean2018}.
Moreover, methods relying on the Lehmann's representation may need to prepare an exponentially large number of excited states \cite{Rungger}.

One important feature of the Green's functions is that they hold information about the single-particle excitation spectrum.
In particular, the (single-particle) spectral function is proportional to the imaginary part of the Fourier transform of the retarded Green's function \cite{Tremblay}.
In this article, we consider the problem of recovering the spectral function from real time measurements of Green's functions.

The previous literature \cite{Wecker, Bauer, Kreula, Jaderberg, Keen, Endo} has always approached the problem essentially the same way.
One measures a Green's function at different times (possibly with multiple runs of a quantum circuit), takes the discrete Fourier transform of the measurements, and associates its peaks with the single-particle excitation energies.
From Gabor's uncertainty principle \cite{Gabor}, the spectral resolution $\delta_f$ is inversely proportional to the maximum measured time $t_{\mathrm{max}}$, that is, $\delta_f \gtrsim 1 / t_{\mathrm{max}} $.
So, it becomes challenging for NISQ computers to produce accurate estimates, as we are in practice restricted to very small time windows due to decoherence errors.

We argue that the aforementioned strategy is not optimal for this problem.
We know there is a finite number of spectral lines, but the Fourier transform does not incorporate this assumption.
The situation is similar to reconstructing a sparse signal in the discrete Fourier basis.
In this setting, compressive sensing methods \cite{CandesWakin}, seeking the sparsest possible representation of the signal in terms of the said basis, recover the signal with fewer measurements than what we would otherwise expect from the Nyquist-Shannon sampling theorem \cite{Shannon}.
However, for the same time window, the standard theory of compressive sensing \cite{CScompact} does not guarantee a better spectral line resolution compared with the discrete Fourier transform.
In essence, we are still discretizing the continuous parameter space as a finite grid, while in general the true spectral lines do not fall into the grid --  this is known as the basis mismatch problem \cite{Chi2011}.

In a breakthrough in signal processing theory, Refs. \cite{SuperRes, OffTheGrid} showed that under certain conditions it is possible to go beyond Gabor's uncertainty limit for the spectral line estimation problem, a result often referred to as \emph{super-resolution}.
These methods work directly on the continuous parameter space, circumventing the limitations imposed by discretization (and so they are also sometimes referred to as ``off-the-grid'' compressive sensing).

Applying the Atomic Norm Minimization technique developed in \cite{AtomicNorm, OffTheGrid, AtomicLSE, MinMax}, we show that we can reach a super-resolution performance for the reconstruction of the spectral function from measurements of Green's functions on quantum computers.
As a demonstration of this technique, we recover the spectral function of a single-impurity model from measurements of the real-time evolution on one of IBM's quantum processors. 
We benchmark our results against the ``naive'' discrete Fourier transform method, showing that only the super-resolution method provides an accurate reconstruction.

\section{A spectral line estimation problem}

The Green's function, denoted here as $G(t)$, is a two-point correlation function that plays a central role in the theory of many-body systems.
Computing it for strongly-correlated systems with many particles is a very demanding task for classical computers \cite{Schuch2009}.
In contrast, if there is an efficient qubit encoding of the fermionic degrees of freedom, and if the corresponding Hamiltonian $\Ham$ is efficiently simulatable  \cite{Childs2018}, we can approximate $G(t)$ by evaluating the expected value of suitable observables on quantum circuits \cite{Wecker, Bauer, Kreula, Jaderberg, Keen, Endo}.
The key routine for these methods is the efficient implementation of the unitary $\exp(-i\Ham t)$.
However, large simulation times will generally require greater circuit depths, which becomes an obstacle in devices with limited coherence times such as the ones in the NISQ computers era.
Therefore, in practice we are limited to computing $G(t)$ for very small values of $t$.

We can write the Green's function as 
\begin{equation}
    G(t) = - i \Theta(t) \sum_{l=1}^s c_l e^{i \omega_l t}.
\label{eq:quantumsignal}
\end{equation}
for some $s \in \N$ and a suitable choice of positive real amplitudes $c_1, \ldots, c_s$ and distinct real energies $\omega_1 \ldots, \omega_s$.
Closely related to the Green's function is the spectral function $A(\omega)$, which can be written as finite sum of weighted Dirac deltas.
Physically, it conveys information about the single-particle excitation spectrum.
From the values of the $c$'s and $\omega$'s we can directly infer the weights and the locations of the poles of the spectral function,
\begin{equation}
    A(\omega) = \sum_{l=1}^s c_l \delta(\omega - \omega_l).
\end{equation}

In the language of signal processing theory, we say that we ``sample the signal'' $G(t)$ at $n$ discrete times, $t_0, t_1, \ldots, t_{\mathrm{max}}$.
Ideally, the samples would constitute an $n$-dimensional vector $\vb{x}^*$ such that $x_j^* = G(t_j)$.
In practice, the sampled signal deviates from the exact Green's function because of approximation errors, hardware noise, and the fact that expectation values of observables are estimated with a finite number of measurements.
Then, we say that we record a noisy signal $\vb{y} = \vb{x}^* + \vb*{\epsilon}$, where $\vb*{\epsilon}$ is additive noise.
The spectral line estimation problem is to recover the amplitudes and energies $\{(c_l, \omega_l)\}_{l=1}^s$ using as few resources as possible.
We mean that we would like to minimize the number of measurements $n$ and, most importantly in the context of NISQ quantum simulation, the maximum sampling time, $t_{\mathrm{max}}$.

In previous proposals on quantum simulation of the Green's function in real time \cite{Wecker, Bauer, Kreula, Jaderberg, Keen, Endo}, the spectral lines are always recovered via the discrete Fourier transform of the signal.
The discretization of the frequency space introduces an error in the frequency estimation, $\delta_f$.
With this approach, both the number of samples, $n$, and the maximum sampling time, $t_{\mathrm{max}}$, scale as    $1 / \delta_f$ (the later being a manifestation of Gabor's uncertainty principle).
Fortunately, as we will see, it is possible to circumvent the $t_{\mathrm{max}} \sim 1 / \delta_f$ scaling  by working directly on the continuous parameter space, entering the so-called super-resolution regime.

\section{Atomic Norm Minimization \label{sec:atomicnorm}}

\begin{figure}[t]
\centering
\includegraphics[width=0.9\linewidth]{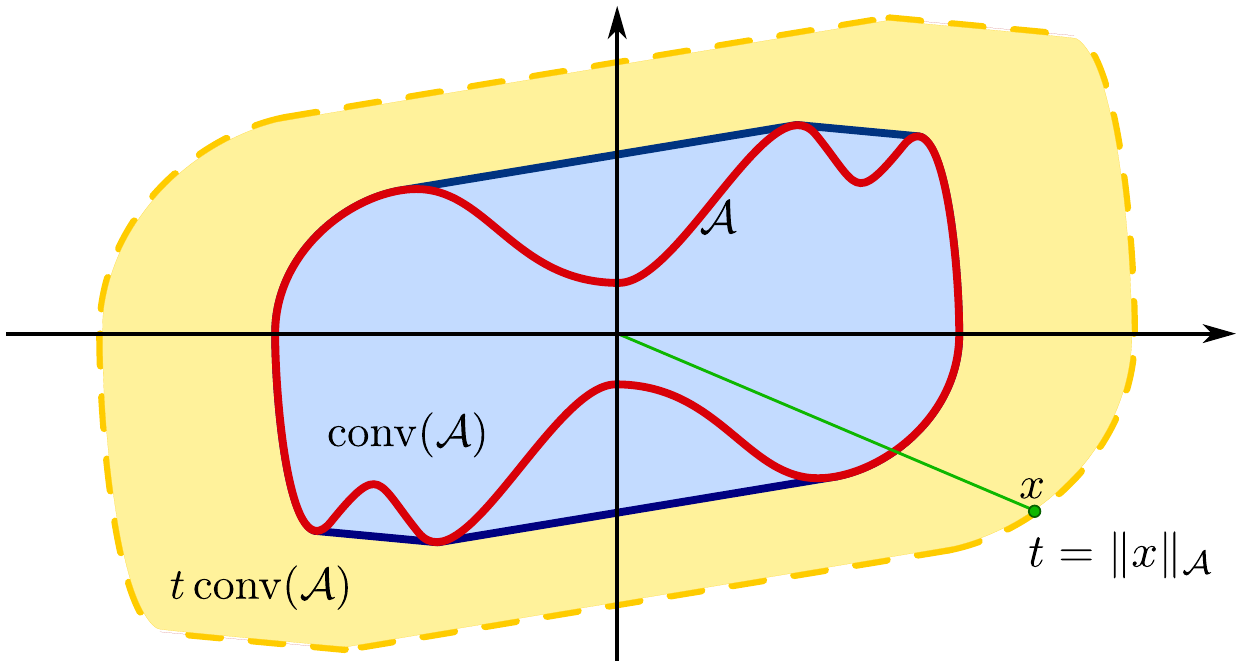}
\caption{
For any non-empty, compact, and centrally-symmetric subset $\A$ of some vector space $V$, we can induce a norm by identifying the $1$-ball with the convex hull of $\A$ \cite{AtomicNorm}.
The atomic norm of any element $\vb{x} \in V$ is the smallest dilation factor $t \geq 0$ such that $\vb{x} \in t \cdot \convA$.
In this Figure (considering a two-dimensional vector space), the red line represents an arbitrary atomic set $\A$, the blue region is $\convA$, and the yellow region is the $t$-ball of $\atom{\cdot}$.}
\label{fig:atomicnorm}
\end{figure}

We now succinctly introduce the super-resolution methods developed in \cite{AtomicNorm, OffTheGrid, AtomicLSE, MinMax}, centered on the concept of the atomic norm.
The key idea is to seek the decomposition of the signal $\vb{y}$ involving the \emph{smallest possible number of atoms}, to be precised in what follows.
For concreteness, choose the sampling times as $t_j = j$ with $j \in \qty{0,\ldots,n-1}$.
Then, define the atoms $\vb{a}(f) \in \C^n$ as
\begin{equation}
a\left(f \right) \vert_j :=  
e^{i 2 \pi f j}, 
\quad j \in \qty{0,\ldots,n-1}.
\end{equation}
The atomic set
\begin{equation}
    \A := \{\vb{a}(f) : f \in [0, 1)\}
    \label{eq:atomicset}
\end{equation}
constitutes the building blocks of our signals.
After proper rescaling (see Supplemental Material \ref{sect:AM_shift}), we can write the noiseless signal, $\vb{x}^*$, as a combination of such atoms.
In fact, there is an infinite number of ways to express $\vb{x}^*$ as a sum of elements of $\A$.
But, in the spirit of Occam's razor, we aim for the simplest description of the signal in terms of this atomic set.
This notion is quantified in terms of the atomic norm $\atom{\cdot}$, defined by identifying its unit ball with the convex hull of $\A$, denoted as $\convA$ \cite{AtomicNorm}.
For any $\vb{x} \in \C^n$, define 
\begin{align}
\atom{ \vb{x} }  
& :=
\inf_{t > 0} 
\big\{ t : \vb{x} \in t \, \convA
\big\} \label{eq:atomnormdef}
\\
& =
 \inf_{\scriptsize \begin{array}{c}c_k \in \C\\ f_k \in [0,1)\end{array}}
\bigg\{
	\sum_k \vert c_k \vert : \vb{x} = \sum_k c_k \, \vb{a}\left(f_k \right)
\bigg\}. 
\end{align}
A decomposition $\sum_k c_k \, \vb{a}(f_k)$ that achieves the infimum is called an atomic decomposition of $\vb{x}$.
See \Cref{fig:atomicnorm} for an illustration of this concept.
While it is not immediately clear from the definition that solving \Cref{eq:atomnormdef} is computationally feasible, Refs. \cite{OffTheGrid, AtomicLSE} show that this can be reduced to a semidefinite program with a reasonably efficient solution.

Having measured a noisy signal $\vb{y}$, we produce an estimate $\hat{\vb{x}}(\vb{y})$ of $\vb{x}^*$ as
\begin{equation}
\hat{\vb{x}}(\vb{y}) = \argmin_{\vb{x} \in \C^n} \qty{ \frac{1}{2} \Vert \vb{y} - \vb{x} \Vert_2^2  + \tau \atom{\vb{x}}},\label{eq:AND}
\end{equation} 
where $\tau > 0$ is a suitably chosen regularization parameter \cite{AtomicLSE}.
Effectively, by solving \Cref{eq:AND} we are denoising the measured signal.
This method can be seen as a generalization of LASSO regression, with the $l_1$ norm replaced by our atomic norm.
Note that, while our atomic set $\A$ is infinite, the $l_1$ norm used in LASSO can be induced by a discrete atomic set -- for example, the canonical basis vectors together with their reflections about the origin.

The relevance of the denoised estimate $\hat{\vb{x}}(\vb{y})$ for the spectral line estimation problem becomes clear in light of Lagrangian dual theory \cite{AtomicLSE}.
For all $f \in [0,1)$, we define the \emph{dual polynomial} $Q$ as
\begin{equation}
    Q_{\vb{y}}(f) := \frac{1}{\tau} 
    \vert
    \left\langle \vb{a}(f) , \vb{y} - \hat{\vb{x}}(\vb{y}) \right\rangle
    \vert, 
    \label{eq:dualpolynomial}
\end{equation}
where $\langle \cdot , \cdot \rangle$ denotes the real inner product.
One can show that $Q$ is upper bounded by $1$.
Moreover, we can write $\hat{\vb{x}}(\vb{y})$ as a sum of atoms $\vb{a}(f)$ only with frequencies for which $Q_{\vb{y}}(f) = 1$.
In other words, the line spectra is identified with the peaks of the dual polynomial, offering a completely different perspective on the problem compared with traditional approaches.

Now the pending question is how close is  $\big\{ \big(\hat{c}_l, \hat{f}_l \big) \big\}_l$ to the true signal $\big\{ \big(c_l, f_l \big) \}_l$ (or, indirectly, how good an estimate of $\vb{x}^*$ is $\hat{\vb{x}}(\vb{y})$).
An answer was provided in Ref. \cite{MinMax} (and later improved by Ref. \cite{TaleOfResolution}), asserting that the reconstruction performance critically depends on the minimal separation between the frequencies, known as the frequency gap, $\Delta_f := \min_{j \neq l} \vert f_j - f_l \vert$, where $\vert \cdot \vert$ is understood as the wraparound distance around the unit circle. 
Then, under some reasonable assumptions on the error, the distribution of the coefficients, and the signal-to-noise ratio \cite{TaleOfResolution}, there is an assignment of $\tau$ in \Cref{eq:AND} for which the estimate is close to the true signal (with high probability), as long as 
\begin{equation}
    t_{\max} \geq 2.5 / \Delta_f.
    \label{eq:theoreticalminimum}
\end{equation}

This behavior is strikingly different from the discrete Fourier transform.
For a given error bound $\delta_f$, with the atomic norm minimization method we only need to sample up to $t_{\mathrm{max}} \sim 1 / \Delta_f$, as opposed to the $t_{\mathrm{max}} \sim 1 / \delta_f$ required with the discrete Fourier transform.
Note that in order to resolve all the spectral lines we need $\delta_f \leq \Delta_f$ and, in practice, we usually want $\delta_f \ll \Delta_f$.
For these cases, the Atomic Norm Minimization will significantly outperform the discrete Fourier transform.

\section{Quantum simulations}

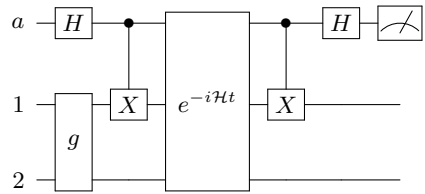
\begin{figure}[t]
\centering
\leavevmode
\Qcircuit @C=0.75em @R=2.em {
\lstick{a}&\gate{H}&\ctrl{1}              &\multigate{2}{e^{-i\Ham t}}  &\ctrl{1}             &\gate{H}& \meter
\\
\lstick{1}                &\multigate{1}{g}&\gate{X}&\ghost{e^{-i\Ham t}}  &\gate{X}&\qw     & \qw
\\
\lstick{2}    &\ghost{g}       &\qw.                  &\ghost{e^{-i\Ham t}}       &\qw.                 &\qw     & \qw }
\caption{Quantum circuit used to compute the Green's function.
It is a variation of the single-qubit interferometry scheme (commonly referred to as the Hadamard test) employed in Reference \cite{Kreula}. $H$ is the Hadamard gate and the $g$ gate denotes the ground state preparation circuit. As shown in Supplemental Material \ref{sec:trottercircuit}, we can use the fact that we are only interested in the expected value of an observable defined on the top qubit to simplify the circuit to only two qubits.}
\label{fig:qcircuit}
\end{figure}

As a demonstration of our technique, We apply the Atomic Norm Minimization (ANM) method to the single-impurity model (also known as the Anderson model) \cite{Georges1996}, which has received wide attention in the context of quantum simulation for its role in dynamical mean-field theory and its relation with the Hubbard model.
Due to quantum hardware limitations, we consider only one fermionic bath site.
For a suitable choice of parameters (corresponding to a fixed point of the dynamical mean-field theory loop \cite{Potthoff}), the single-impurity Hamiltonian can be encoded into the Hilbert space of three qubits with the Bravyi-Kitaev transformation as (cf. Supplemental Material \ref{sec:BK})
\begin{equation}
    \Ham  = Z_1 Z_2 + 0.75 X_2 + 0.37 \qty(1+Z_a) X_1.
    \label{eq:1_site_hamiltonian}
\end{equation} 
Letting $\mathcal{C}$ be the unitary encoded in the quantum circuit of Figure \ref{fig:qcircuit}, the impurity's Green's function can be expressed as
\begin{equation}
    G_{\mathrm{imp}}(t) =  -i\Theta(t) \expval{\mathcal{C}^{\dg} Z_a \mathcal{C}} ,
    \label{eq:G(t)_1site}
\end{equation}
where the expectation value is over the all-zero state.

For several times $t$, we evaluate $G_{\mathrm{imp}}(t)$ with the circuit represented in Figure \ref{fig:qcircuit} using IBM's \texttt{ibmq\_manila} device, following Ref. \cite{Kreula}.
The details of our implementation, including a strategy to exploit the symmetries of the spectral function, are explained in Supplemental Materials \ref{sect:quantum_circuit}-\ref{sect:error_mitigation}.
But the core routine of the computation remains the time evolution of the system's ground state.
For this, we approximate the operator $\exp(-i \Ham t)$ with a second order Suzuki expansion \cite{Suzuki1991, Childs2018}, which we simplify down to a circuit that only contains two two-qubit gates.
In our simulations, we assume that we know \emph{a priori} the ground state, which can be determined with a simple classical calculation and prepared with a short sub-circuit.
In a scenario where the ground state is unknown, we could prepare it, for example, using variational quantum algorithms \cite{VQE}.
We opt not to do this here, as our focus is in the performance of the ANM method.
Finally, we resort to some standard error mitigation techniques to improve the precision of the measurements.

With the measured values of  $G_{\mathrm{imp}}$, we reconstruct the spectral function with two distinct methods:

\emph{Discrete Fourier transform} (DFT). We apply a standard DFT-based method.
We use zero padding \cite{MDFT07} to help determine the location of the peaks, noting that this does not actually increase the resolution.
We estimate the amplitudes and energies $\big\{ \big(\hat{c}_l, \hat{\omega}_l \big) \big\}_l$ by matching each observed peak with the expected theoretical form of a \texttt{sinc} function. To discern the contribution of the different peaks, we iteratively remove the observed peaks from the spectrum, going from the peak with the highest to the lowest amplitude.

\emph{Atomic Norm Minimization} (ANM).
After proper rescaling of time (cf. \Cref{sect:AM_shift}), we denoise the observed signal with ANM, solving \Cref{eq:AND}.
The frequencies $\big\{ \hat{f}_l \big\}_l$ are identified with the peaks of the dual polynomial $Q(f)$ of the denoised signal, and are then converted to the energies $\big\{ \hat{\omega}_l \big\}_l$.
Finally, the amplitudes $\big\{ \hat{c}_l \big\}_l$ are retrieved via least squares minimization.

\vspace{1.5em}

\begin{figure}[t]
\centering
\includegraphics[width=\linewidth]{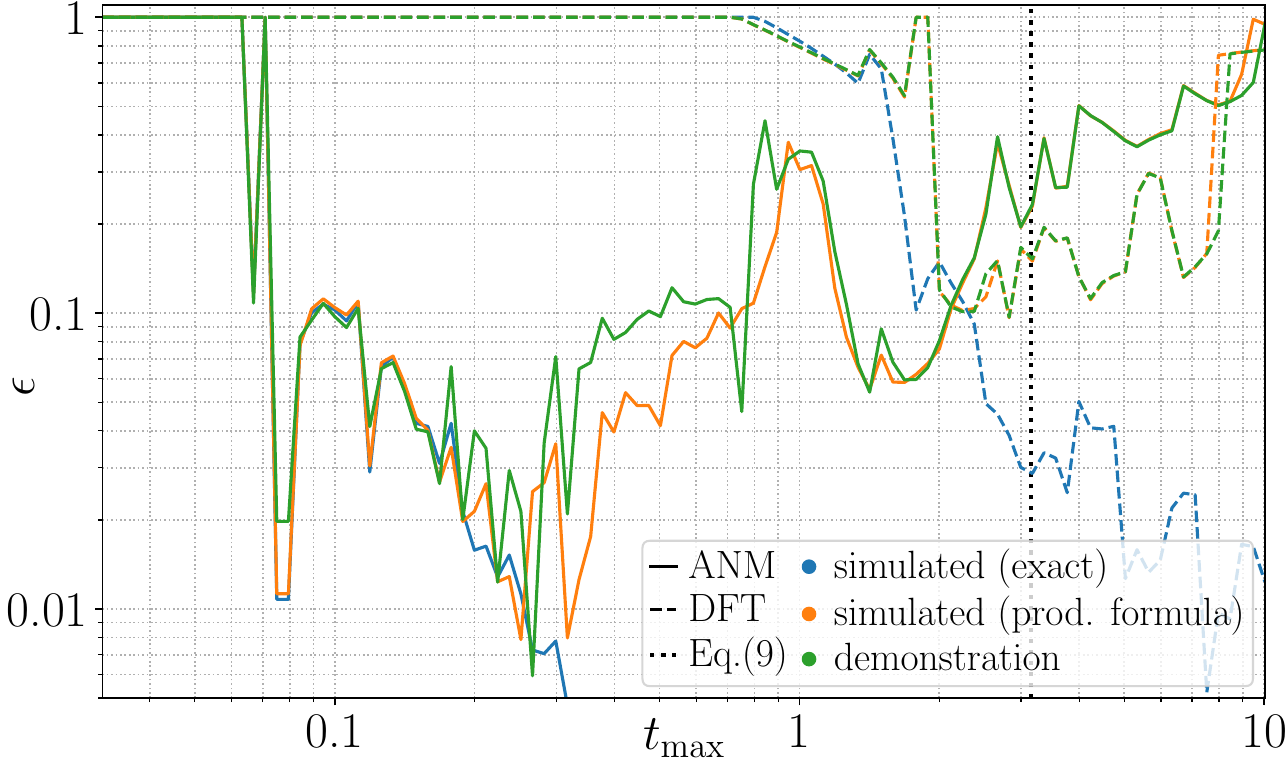}
\caption{
Spectrum reconstruction error \eqref{eq:reconstruction_error} as a function of $t_{\mathrm{max}}$ for the ANM and the DFT methods  -- full and dashed lines, respectively.
The blue lines correspond to a perfect simulation of $\exp(-i\Ham t)$ and exact evaluation of the observables. 
The orange lines also originate from simulated data, but include only the error from the second order Suzuki formula (no circuit or observable estimation errors).
Finally, the green lines are obtained from the demonstration on IBM's \texttt{ibmq\_manila} device (see \Cref{sect:hardware_configuration}).
For the demonstration, we took $10^5$ shots of the quantum circuit for each value of $t$.
For $t_{\mathrm{max}} =0.27$ the ANM reaches a reconstruction error of $0.5 \%$.
Nevertheless, for larger values of $t_{\mathrm{max}}$ the ANM begins to worsen because the product formula that we use starts to deviate from the exact time evolution.
Independently of the value of $t_{\mathrm{max}}$, the DFT never achieves a reconstruction error below $\sim 10\%$, even in a noiseless scenario. 
The black dotted vertical line signals the value of $t_{\mathrm{max}}$ past which the ANM theory guarantees a successful reconstruction (cf. Equation \eqref{eq:theoreticalminimum}).
We observe that in practice we need much shorter time windows.
}
\label{fig:1site_DFTANM}
\end{figure}

In Figure \ref{fig:1site_DFTANM}, we compare the performance of the DFT and the ANM methods for various values of $t_{\mathrm{max}}$.
For this purpose, we define the reconstruction error to be
\begin{equation}
    \epsilon := \frac{\sum_l  |c_l| \cdot \vert \omega_l - \hat{\omega}_l \vert + \vert c_l - \hat{c}_l \vert}{\sum_j |c_j| \cdot |\omega_j| + |c_j|},
    \label{eq:reconstruction_error}
\end{equation}
thereby encapsulating the error in the estimation of both the frequencies and the amplitudes.

For a perfect reconstruction, we would expect four poles, $\{( c_l, \omega_l)\}_l = \qty{(0.525, \pm 0.548), (0.475, \pm 3.042)}$.
The DFT never reaches a reconstruction error below $9.7\%$, obtaining the poles $\qty{(0.528, \pm 0.265), (0.460, \pm 2.836)}$.
This is because when the time window becomes large enough to allow a precise spectral location, the low order Suzuki expansion has already deviated significantly from the exact time evolution operator.
In contrast, the error of the ANM method decreases sharply for small values of time, reaching $\epsilon = 0.5\%$ for $t_{\mathrm{max}} = 0.27$, and obtaining $\{( c_l, \omega_l)\}_l =  \qty{(0.524, \pm 0.562), (0.475, \pm 3.025)}$.
After that value of $t_{\mathrm{max}}$ the ANM error also grows due to the time evolution approximation errors.
This order of magnitude difference between the results of the ANM and the DFT constitutes evidence that the former can offer an important advantage in reconstructing spectral functions on real noisy quantum devices.

\section{Discussion \label{sec:conclusion}}

We have proposed the Atomic Norm Minimization, a super-resolution technique well-known in signal processing theory, as a tool for studying the single-particle excitation spectrum of quantum systems with quantum computers.
In previous works, the reconstruction of the spectral function from measurements of the Green's function at discrete times was approached with the discrete Fourier transform.
In this case, by Gabor's uncertainty principle, the resolution of the spectral lines scales with the inverse of the maximum sampling time.
In contrast, with the Atomic Norm Minimization this resolution scales with the inverse of the spectral gap, which in practice greatly reduces the depth requirements of the quantum circuits used to compute the Green's function.
Applying these techniques, we were able to reconstruct the spectral function of the single-impurity model with one bath site using IBM's \texttt{ibmq\_manila} with an error one order of magnitude smaller than with standard signal reconstruction methods (based on the discrete Fourier transform).

With super-resolution techniques, we can reach tolerable reconstruction errors by sampling the Green's function up to a $t_{\rm max}$ that is one order of magnitude lower than what is required with more standard methods.
The less stringent requirements allow using lower order product formulas that deviate from the exact time-evolution quicker, but also require fewer quantum gates to implement.
With fewer gates, the simulations are less affected by the high error rates associated with currently available NISQ hardware.
Ultimately, these ideas may enable the simulation of previously unexplored systems, and possibly showcase quantum advantage in a nearer future.

As far as we are aware, ours constituted the most accurate reconstruction of the spectral function of the single-impurity model with one bath site based on measurements of the real-time Green's function using real quantum hardware.
We note that Ref.~\cite{Keen} succeeded in this task only for the case $V=0$, which is essentially a single site problem.
The other experimental calculations of the spectral function found in the literature \cite{Endo, Rungger,  Zhu} were based on the Lehmann's representation and resorted to variational quantum eigensolvers.
Compared with these variational methods, the computational scaling of our approach is better characterized, with proven efficiency in the limit of many particles, as long as the spectral gap remains sufficiently large.
Our work stands from most publications in the field for being a successful quantum simulation that is not intrinsically based on variational algorithms.
Moreover, unlike previous approaches, ours does not require knowing in advance the number of poles in the spectral function. 

We have observed that we can reconstruct the spectral function with time windows significantly shorter than what is guaranteed by the Atomic Norm Minimization theory (Equation \ref{eq:theoreticalminimum}).
This can be partially explained by the theory of super-resolution of positive sources \cite{Chi_DaCosta_2019, DaCosta_Chi_2021}, which considers the restricted version of the problem where the coefficients $c_l$ in Equation \eqref{eq:quantumsignal} are real positive numbers.
Up to a global phase factor, the spectral function fits this constraint.
In this case, one can show that we can perfectly reconstruct the signal from any $s+1$ \emph{noiseless} samples, independently of the spectral gap.
As far as we know, these results have not been extended to a noisy setting.
Therefore, although the guarantees that we present are fully rigorous, they may be overly conservative and it may be possible to tighten them significantly.

The Theorems of \cite{MinMax, TaleOfResolution} about the performance of the Atomic Norm Minimization assumed the noise to be composed of independently identically distributed complex Gaussian random variables, which is not the case in our demonstration.
Despite this, the reconstruction was successful, reaching an error of $0.5\%$.
An interesting question is how close is the signal that we measure to this assumption, or how can the theoretical guarantees be generalized to other noise models.

To the best of our knowledge, this is the first time that the Atomic Norm Minimization is discussed in the context of quantum computing.
While we have considered it for the study of the single-particle excitation spectrum, our methods may benefit other problems.
Examples of possible applications include probing many-body localization \cite{ManyBodLoc}, learning the structure of entanglement Hamiltonians \cite{EntanglementHamiltonian}, and studying Floquet systems \cite{Floquet}.
In general, super-resolution techniques can be used when the signal read from the quantum processor can be framed as a sum of a few atoms from a possibly infinite atomic set (not just the Fourier basis).
Hopefully, future works will unveil the full potential of the Atomic Norm Minimization for quantum computing.

\begin{acknowledgments}

We would like to thank M\'{a}rio Figueiredo for introducing us to off-the-grid compressive sensing ideas and Maxime da Costa for sharing his expertise on the atomic norm.
We thank Yasser Omar for supervision and reviewing the manuscript.
We would also like to thank Jo\~{a}o Moutinho, Miguel Murça, Pedro Ribeiro, Sagar Pratapsi, and Stefano Gherardini for fruitful discussions.

The authors thank the support from FCT – Funda\c{c}\~{a}o para a Ci\^{e}ncia e a Tecnologia (Portugal), namely through projects UIDB/50008/2020 and UIDB/04540/2020, as well as projects QuantHEP and HQCC supported by the EU H2020 QuantERA ERA-NET Cofund in Quantum
Technologies and by FCT (QuantERA/0001/2019 and QuantERA/004/2021, respectively). DC and DM acknowledge the support from FCT through scholarships UI/BD/152301/2021 and 2020.04677.BD, respectively.

\end{acknowledgments}

\appendix

\section{Rescaling the signal \label{sect:AM_shift}}

In order to implement Atomic Norm Minimization, we assume a signal of the form
\begin{align}
    x_j = \sum_l c_l e^{i 2\pi f_l j}\label{eq:x_j},
\end{align}
where $f_l \in [0,1)$ and $j \in \qty{0,\ldots, n-1}$. 
To cast the signal in this form, we may need to perform energy and time shifts.
For this, we require an estimate of the energy range $[\omega_a, \omega_b]$ where the peaks of the spectral function are found, and an estimate of the smallest spectral gap. We know that, for a Hamiltonian
\begin{align}
    \Ham = \sum_k h_k (P_{k, 1}\otimes P_{k, 2}\otimes \cdots),\quad \text{with }P_{k,j} \in \qty{I,X,Y,Z},
\end{align}
its minimum and maximum eigenenergies are bounded by
\begin{align}
    E_{\min{}} \geq -\sum_k |h_k|,\qquad E_{\max{}} \leq \sum_k |h_k|.
\end{align}
Consequently, the highest peak (in absolute value) for the spectral function is bounded by $E_{\max{}} - E_{\min{}}$. If no better estimate of the energy range is known, then we may set
\begin{align}
    \omega_a =  - 2 \sum_k |h_k|, \qquad \omega_b = +2 \sum_k |h_k|.
\end{align}
Assuming that the spectral function has at least $K$ non-zero peaks, then we may set an upper bound $\Delta_\omega$ for the spectral gap as
\begin{align}
    \Delta_\omega = \frac{\omega_b - \omega_a}{K-1}.
\end{align}
If more information is known about the quantum system at hand, the energy range $[\omega_a, \omega_b]$ may be tightened, and the spectral gap estimate $\Delta_\omega$ lowered from these general estimates.
Due to the wraparound nature of the energy/frequency range in the context of Atomic Norm Minimization, we must first pad the energy range to ensure the wraparound region does not introduce an artificially small energy gap. 
As a result, the effective energy range considered is
\begin{align}
    \qty[\omega_a - \frac{\Delta_\omega}{2}, \omega_b + \frac{\Delta_\omega}{2}]
\end{align}
and so the effective frequency range is 
\begin{equation}
    \Omega_{\max} := \frac{1}{2\pi}\qty(\omega_b - \omega_a + \Delta_\omega).
\end{equation}

Now suppose that we take $n$ samples of $\Gu(t)$ at times
\begin{gather}
    T = \qty{t_j}_{j=0}^{n-1}, \\
    t_j := t_0 + \frac{j}{\Omega_{\max{}}},
\end{gather}
obtaining the signal
\begin{equation}
    x_j^{\mathrm{real}} = \Gu(t_j).
\end{equation}
Then, to reach the desired form for Atomic Norm Minimization, we perform the mappings
\begin{align}
    \qty{t_j}_{j=0}^{n-1} &\longmapsto \qty{0,\ldots, n-1},\\
    \qty[\omega_a - \frac{\Delta_\omega}{2}, \omega_b + \frac{\Delta_\omega}{2}] &\longmapsto [0, 2\pi],
\end{align}
for the time and energy variables, respectively. The latter mapping ensures that the frequencies are contained in $[0, 1)$, since $\omega = 2\pi f$.
Let
\begin{equation}
    \phi := \frac{1}{2\pi\Omega_{\max{}}}\qty(\omega_a - \frac{\Delta_\omega}{2}).\label{eq:phi_ANM}
\end{equation}
$\phi$ represents a phase shift. 
For example, $\phi=0$ indicates that our energy range already starts at zero (so no shift is required), while $\phi=-1/2$ indicates an originally symmetric energy range, of the form $[-\omega, \omega]$.
Applying the correction
\begin{equation}
    x_j = x_j^{\mathrm{real}} e^{-i 2 \pi \phi \Omega_{\max} t_j},
\end{equation}
the rescaled signal $x_j$ satisfies \Cref{eq:x_j}, as desired. Applying Atomic Norm Minimization to the signal $x_j$, we obtain the estimate $\qty{(\hat{c}_l, \hat{f}_l)}_{l=1}^s$. By undoing the rescaling with
\begin{align}
    c_l^{\mathrm{real}} &= \hat{c}_l e^{-i 2 \pi \hat{f}_l \Omega_{\max} t_0},\\
    \omega_l^{\mathrm{real}} &= 2\pi f_l^{\mathrm{real}} = 2\pi\Omega_{\max{}}(\hat{f}_l + \phi),
\end{align}
we obtain the spectrum $\qty{(c_l^{\mathrm{real}}, \omega_l^{\mathrm{real}})}_{l=1}^s$ associated with our original signal $x_j^{\mathrm{real}}$.

For the discrete Fourier transform, the frequencies are considered to belong to the $[-1/2, 1/2]$ range, instead of the $[0, 1)$. 
We use the previous approach, but now the phase shift
\begin{equation}
    \phi_{\mathrm{DFT}} = \phi + \frac{1}{2}
\end{equation}
increases by $1/2$, when compared with the Atomic Norm Minimization phase shift of \Cref{eq:phi_ANM}. Note that, in this case, the spectral gap $\Delta_\omega$ no longer plays the same role. Nonetheless, it may still be set to a nonzero value for the rescaling, to improve resolution for the energies near $\omega_a$ and $\omega_b$.

\section{Derivation of the effective Hamiltonian \label{sec:BK}}

A general impurity Hamiltonian can be written as  
\begin{align}
    \Ham &= \Ham_{\text{imp}} + \Ham_{\text{bath}} + \Ham_{\text{mix}}\label{eq:SIAM_general}\\
    \Ham_{\text{imp}} &= \sum_\alpha (\epsilon_\alpha - \mu) c_\alpha^\dag c_\alpha + 
    \sum_{\alpha\beta\gamma\delta} U_{\alpha\beta\gamma\delta} \, c_\alpha^\dag c_\beta^\dag c_\gamma c_\delta\\
    \Ham_{\text{mix}} &= \sum_{\alpha i} V_{\alpha i} \, c_\alpha^\dag c_i + \text{h.c.}\\
    \Ham_{\text{bath}} &= \sum_i \epsilon_i \, c_i^\dag c_i \;,\label{eq:H_bath}
\end{align}
where the Greek and Latin indices correspond to the impurity and bath fermionic degrees of freedom, respectively, covering both the different sites and their spins. 
$\mu$ is the chemical potential, $\epsilon_{\alpha}$/$\epsilon_i$ the on-site energies of the impurity/bath, $U_{\alpha\beta\gamma\delta}$ the electron interaction energies, and $V_{\alpha i}$ the hopping elements between the impurity and the bath.

We consider the single-impurity model for one bath site, with $\epsilon_{\alpha} = \epsilon_i = 0$, $\mu = U/2$. There are two spin values associated to the impurity and bath site.
Applying the Bravyi-Kitaev transform we get
\begin{gather}
    c_{0,\uparrow} = \frac{X_1 + iY_1}{2}X_2X_4 \label{eq:c_BK}\\
    \Ham_{BK} = \frac{U}{4} Z_1Z_3 + \frac{V}{2}(X_1 - X_1 Z_2 + X_3 - Z_2X_3 Z_4),
\end{gather}
where $0,\uparrow$ refers to the spin-up case in the impurity site.
Thanks to particle number conservation and particle-hole symmetry, for the ground state, the second and fourth qubits always read $1$ and $0$, respectively.
Consequently, for the state $\ket{\text{EX}} := X_1 X_2 X_4\ket{\text{GS}}$ the second and fourth qubits must always read $0$ and $1$, respectively, as they are simply the negated values associated with the ground state.
The Green's function of interest can be written as
\begin{equation}
    G_{0,\uparrow}(t) = -i\Theta(t)\ev{\{\mathcal{U}^\dagger c_{0,\uparrow} \mathcal U,c_{0,\uparrow}^\dagger\}}{\text{GS}},
\end{equation}

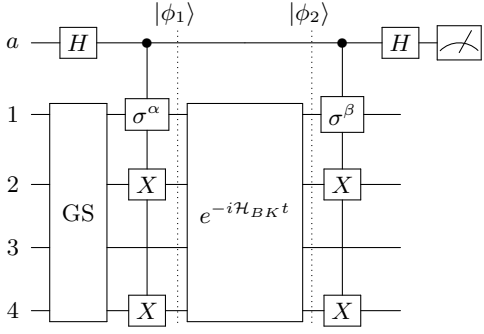
\begin{figure}[t]
\centering
\leavevmode
\Qcircuit @C=0.75em @R=1.5em {
& & \raisebox{-2.5em}{\qquad\;$\ket{\phi_1}$} &\raisebox{-2.5em}{\qquad\qquad\quad\;$\ket{\phi_2}$} & & & &\\
\lstick{a}&\gate{H}&\ctrl{1}\ar@{.}[]+<1.25em,0.5em>;[dddd]+<1.25em,-0.5em>            &\qw  &\ctrl{1} \ar@{.}[]+<-1.25em,0.5em>;[dddd]+<-1.25em,-0.5em>            &\gate{H}& \meter
\\
\lstick{1}                &\multigate{3}{\text{GS}}&\gate{\sigma^{\alpha}}&\multigate{3}{e^{-i\Ham_{BK} t}}  &\gate{\sigma^{\beta}}&\qw     &
\\
\lstick{2}    &\ghost{\text{GS}}       &\gate{X}\qwx                  &\ghost{e^{-i\Ham_{BK} t}}       &\gate{X}\qwx                 &\qw     &\\
\lstick{3}                &\ghost{\text{GS}}&\qw &\ghost{e^{-i\Ham_{BK} t}}  &\qw &\qw     &
\\
\lstick{4}    &\ghost{\text{GS}}       &\gate{X}\qwx[-2]                  &\ghost{e^{-i\Ham_{BK} t}}       &\gate{X}\qwx[-2]                 &\qw     &}
\caption{Original single-qubit interferometry scheme used to compute the Green's function, before the simplifications that led to the one in \Cref{fig:qcircuit}.
$\sigma^{\alpha / \beta}$ means an $X$ or a $Y$ Pauli gate, $H$ is the Hadamard gate, and the $\text{GS}$ operator denotes the ground state preparation circuit.}
\label{fig:qcircuit_BK}
\end{figure}

For the particular Hamiltonian we are considering, the Green's function is symmetric, so we only need to consider the case $\sigma^\alpha=\sigma^\beta=X$ in \Cref{fig:qcircuit_BK} (see Supplemental Material \ref{sect:quantum_circuit}, where this result is derived, and the resulting \Cref{eq:G(t)_1site}). Looking at \Cref{fig:qcircuit_BK}, we may conclude that
\begin{align}
    \ket{\phi_1} &= \frac{1}{\sqrt{2}}\ket{0}_a\ket{\text{GS}}_{1234} + \frac{1}{\sqrt{2}}\ket{1}_a \ket{\text{EX}}_{1234}\\
    &= \frac{1}{\sqrt{2}}\ket{0}_a\ket{g}_{13}\ket{10}_{24} + \frac{1}{\sqrt{2}}\ket{1}_a \ket{e}_{13}\ket{01}_{24},
\end{align}
with $\ket{g},\ket{e}$ some quantum states in the $1,3$ subspace.
Since $\Ham_{BK}$ either acts as the identity or $Z$ in qubits 2 and 4, then we have, from \Cref{fig:qcircuit_BK},
\begin{equation}
    \ket{\phi_2} = \frac{1}{\sqrt{2}}\ket{0}_a\ket{g_f}_{13}\ket{10}_{24} + \frac{1}{\sqrt{2}}\ket{1}_a \ket{e_f}_{13}\ket{01}_{24},
\end{equation}
with $\ket{g_f},\ket{e_f}$ some quantum states in the $1,3$ subspace. We observe that the values of qubits 2 and 4 are preserved throughout the Hamiltonian evolution.
Consequently, these qubits may be tapered off.
Renaming the third qubit as the second one, we reach two distinct Hamiltonians,
\begin{align}
    \Ham_{BK, GS} &= \frac{U}{4} Z_1Z_2 + V (X_1 + X_2)\\
    \Ham_{BK, EX} &= \frac{U}{4} Z_1Z_2 + V X_2,
\end{align}
for the time evolutions of $\ket{\text{GS}}$ and $\ket{\text{EX}}$, respectively. These are the Hamiltonians such that
\begin{align}
    \ket{g_f}_{12} &= e^{-i\Ham_{BK,GS} t}\ket{g}_{12}\\
    \ket{e_f}_{12} &= e^{-i\Ham_{BK,EX} t}\ket{e}_{12}.
\end{align}

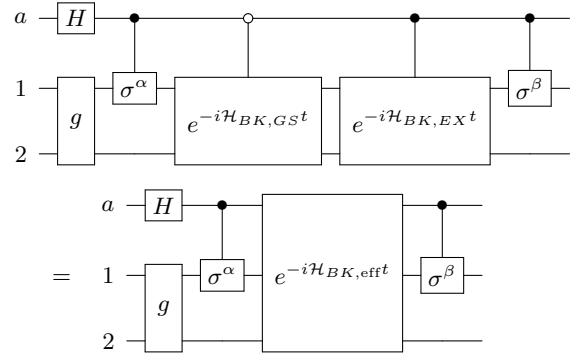
\begin{figure}[t]
\centering
\leavevmode
$$\Qcircuit @C=0.75em @R=1.5em {
\lstick{a}&\gate{H}&\ctrl{1}              &\ctrlo{1}  &\ctrl{1}             &\ctrl{1} & \qw
\\
\lstick{1}                &\multigate{1}{g}&\gate{\sigma^{\alpha}}&\multigate{1}{e^{-i\Ham_{BK,GS} t}}  &\multigate{1}{e^{-i\Ham_{BK,EX} t}}&\gate{\sigma^{\beta}} & \qw
\\
\lstick{2}    &\ghost{g}       &\qw.                  &\ghost{e^{-i\Ham_{BK,GS} t}} &\ghost{e^{-i\Ham_{BK,EX} t}}      &\qw & \qw}$$
$$\Qcircuit @C=0.75em @R=1.5em {
\lstick{a}&\gate{H}&\ctrl{1}              &\multigate{2}{e^{-i\Ham_{BK,\text{eff}} t}}             &\ctrl{1} & \qw
\\
\lstick{=\quad 1}                &\multigate{1}{g}&\gate{\sigma^{\alpha}}&\ghost{e^{-i\Ham_{BK,\text{eff}} t}}  &\gate{\sigma^{\beta}} & \qw
\\
\lstick{2}    &\ghost{g}       &\qw.                  &\ghost{e^{-i\Ham_{BK,\text{eff}} t}}      &\qw & \qw}$$
\caption{The circuit in \Cref{fig:qcircuit_BK} can be partly reduced to this circuit. The controlled evolutions of $\Ham_{BK,GS}$ and $\Ham_{BK,EX}$ can be further simplified into the evolution of an effective Hamiltonian $\Ham_{BK,\text{eff}}$. The gate $g$ denotes the preparation of the state $\ket{g}$, which is the reduced form of $\ket{\text{GS}}$, once the original qubits 2 and 4 have been tapered off.}
\label{fig:qcircuit_eff}
\end{figure}

As a result of this simplification, the simulation procedure can be performed using just 3 qubits, instead of the initial 5 (when counting with the extra ancilla qubit). Unfortunately, this requires evolving the Hamiltonian $\Ham_{BK, GS}$ when the ancilla qubit is $\ket{0}$ and $\Ham_{BK, EX}$ when the ancilla qubit is $\ket{1}$, thereby requiring the evolution operator to be conditionally performed (cf. \Cref{fig:qcircuit}). 
At first glance, it might seem like this could significantly worsen the simulation performance. However, an equivalent evolution to the one intended can be obtained by simulating the effective Hamiltonian
\begin{align}
    \Ham_{BK, \text{eff}} &= \frac{U}{4} Z_1Z_2 + V X_2 + V \qty(\frac{1+Z_a}{2}) X_1
\end{align}
using not only the 2 site qubits, but also the ancilla qubit $a$. 
Within this setting, without the original qubits 2 and 4, the annihilation operator $c_{0,\uparrow}$ can be implemented simply as $\frac{X_1 + iY_1}{2}$.

For our simulations, we choose $U = 4., V = 0.745$, corresponding to a fixed point of the dynamical mean-field theory loop \cite{Potthoff}.

\section{Computing the Green's function \label{sect:quantum_circuit}}

For the circuit in \Cref{fig:qcircuit}, using Pauli matrices $\sigma^{\alpha/\beta}$, if we measure the ancilla qubit in the $Z$ and $-Y$ basis, respectively, we obtain the expectation values
\begin{align}
    E_Z^{\alpha\beta}&  = \Re\ev{{\mathcal U}^\dagger \sigma^\beta {\mathcal U} \sigma^\alpha}, \quad E_{-Y}^{\alpha\beta} = \Im\ev{{\mathcal U}^\dagger \sigma^\beta {\mathcal U} \sigma^\alpha},
\end{align}
corresponding to the real and imaginary parts of $\ev{{\mathcal U}^\dagger \sigma^\beta {\mathcal U} \sigma^\alpha}$.

The Green's function is
\begin{align}
    \Gu(t) &= -i\Theta(t)\ev{\qty{c_\mu(t),c_\mu^\dagger(0)}}{\text{GS}}\\
    &= -i\Theta(t)\ev{\qty{\mathcal{U}^\dagger c_\mu(0) \mathcal U,c_\mu^\dagger(0)}}{\text{GS}},
\end{align}
where $\qty{\cdot, \cdot}$ is the anticommutator. Without loss of generality, we consider that $c_\mu = \frac{X+iY}{2}$ in this section. Its conclusions can be trivially extended to cases where $c_\mu = \frac{X+iY}{2}P$, with $P$ some combination of Pauli operators, as in \Cref{eq:c_BK}.

The impurity's Green's function is then given by
\begin{multline}
    \Gu(t) = \frac{-i\Theta(t)}{4} \Big[ \ev{{\mathcal U}^\dagger X {\mathcal U} X} + \ev{X{\mathcal U}^\dagger X {\mathcal U}}\\
    -i\ev{{\mathcal U}^\dagger X {\mathcal U} Y} - i\ev{Y{\mathcal U}^\dagger X {\mathcal U}}\\
    +i\ev{{\mathcal U}^\dagger Y {\mathcal U} X} + i\ev{X{\mathcal U}^\dagger Y {\mathcal U}}\\
    +\ev{{\mathcal U}^\dagger Y {\mathcal U} Y} + \ev{Y{\mathcal U}^\dagger Y {\mathcal U}}\Big].
    \label{eq:Gmutrain}
\end{multline}
For any Hermitian operator $A$, and eigenstates $\ket{l}$ of $H$, we can see that
\begin{align}
    \ev{{\mathcal U}^\dagger A {\mathcal U} A}{\text{GS}} &= e^{iE_0 t}\ev{A {\mathcal U} A}{\text{GS}}\\
    &= e^{iE_0 t}\ev{{\mathcal U}}{\psi},\\
    \text{with }&\ket \psi = A\ket{\text{GS}}.
\end{align}
Writing $\ket \psi = \sum_l a_l \ket{l}$ using the $\ket l$ eigenstates then leads to
\begin{align}
    \ev{{\mathcal U}^\dagger A {\mathcal U} A}{\text{GS}}&=e^{iE_0 t}\sum_l |a_l|^2 \ev{{\mathcal U}}{l}\\
    &=\sum_l |a_l|^2 e^{-i\omega_l t},\\
    \text{with }&\omega_l:=E_l-E_0 \geq 0.\label{eq:G_XX}
\end{align}
Therefore $\ev{{\mathcal U}^\dagger X {\mathcal U} X}$ and $\ev{{\mathcal U}^\dagger Y {\mathcal U} Y}$ always have real, positive coefficients. Similarly, we may show that
\begin{align}
    \ev{{\mathcal U}^\dagger Y {\mathcal U} X}{\text{GS}} &= -\ev{{\mathcal U}^\dagger X {\mathcal U} Y}{\text{GS}}\\
    &= \sum_l i|a_l|^2 e^{-i\omega_l t} \ev{Z}{l}.\label{eq:G_XY}
\end{align}
Since the expectation value of $Z$ is always real,  these components' coefficients are always imaginary (but possibly negative, if $\ev{Z}$ is negative); and opposite. 
Then, we only need to compute one of these terms for \Cref{eq:Gmutrain}.
Any general impurity Hamiltonian (\ref{eq:SIAM_general}) is particle number preserving, and can be decomposed into $(n+1)$ invariant subspaces $\mathcal S_k$, each associated with an occupation number $k$ from zero to $n$. Therefore, $X$ transforms the ground state $\ket{\text{GS}}\in \mathcal S_g$ into some state $\ket{E_0}:=\alpha_{g-1}\ket{e_{g-1}}+\alpha_{g+1}\ket{e_{g+1}}$, with $\ket{e_k} \in \mathcal S_{k}$. Applying $Z$ leads to $\ket{E_1} := \alpha_{g-1}\ket{e_{g-1}}-\alpha_{g+1}\ket{e_{g+1}}$. We have $\ev{\mathcal U}{E_0}=\ev{\mathcal U}{E_1}$, since $\Ham$ acts independently on the two subspaces $\mathcal S_{g\pm 1}$. As $Y=-iZX$, we conclude that
\begin{equation}
 \ev{{\mathcal U}^\dagger X {\mathcal U} X}{\text{GS}} = \ev{{\mathcal U}^\dagger Y {\mathcal U} Y}{\text{GS}},
 \label{eq:X=Y}
\end{equation}
for Hamiltonians as in \Cref{eq:SIAM_general}.
For the case considered in \Cref{eq:1_site_hamiltonian} of one fermionic bath site, we may go one step further: thanks to particle-hole symmetry, $\ev{Z}{l}=0$ for all $l$, and so cross-term components are zero. These simplifications result in \Cref{eq:G(t)_1site}.

Using \Cref{eq:G_XX,eq:G_XY}, the Green's function can be written as
\begin{align}
    \Gu(t) &= -i \Theta(t) \sum_l |a_l|^2 \Big(\Re\qty{e^{-i\omega_l t}}\notag\\
    &\qquad\qquad\qquad\quad +i\ev{Z}{l}\Re\qty{ie^{-i\omega_l t}}\Big),\\
    &= -i \Theta(t) \sum_l |a_l|^2 \qty(\cos(\omega_l t) + i\ev{Z}{l}\sin(\omega_l t)), \qquad\\
    &= -i \Theta(t) \sum_l\frac{|a_l|^2}{2} \Big[(1+\ev{Z}{l})e^{i\omega_l t}\notag\\
    &\qquad\qquad\qquad\qquad\quad +(1-\ev{Z}{l})e^{-i\omega_l t}\Big].
\end{align}
Note that the Green's function without the $\Theta(t)$ component is symmetric if $\ev{Z}{l} = 0$.
In general, it is simpler to work with
\begin{align}
    \tilde G_\mu(t) &:= \sum_l\frac{|a_l|^2}{2} \qty[(1+\ev{Z}{l})e^{i\omega_l t}+(1-\ev{Z}{l})e^{-i\omega_l t}],
\end{align}
since it holds information at $t<0$ (for a non-symmetric $G(t)$).
This is useful as we may also time-evolve the system in the negative direction, but are limited to evolution times with short absolute values.
The identical peaks and amplitudes of $\tilde G_\mu(t)$ and $\Gu(t)$ then allow us to fully reconstruct $\Gu(t)$. In the case of a symmetric $\Gu(t)$, to avoid having to find two peaks for every $\omega_j$ value, we may compute instead
\begin{align}
    \tilde G_{\text{sym}}(t) &= \sum_l |a_l|^2 e^{i\omega_l t}\\
    &= \ev{{\mathcal U}^\dagger X {\mathcal U} X}^\dagger\\
    &= E_Z^{xx}-iE_{-Y}^{xx},\label{eq:G_sym_xx}
\end{align}
from which $\Gu(t)$ can be obtained.

\section{Ground state preparation circuit}\label{sect:ground_state}

Using the reduced Hamiltonian $\Ham_{BK, \text{eff}}$ for the single-impurity model with one bath site, the ground state is of the form
\begin{align}
    \ket{g} = \cos(\frac{\theta}{2})\qty(\frac{\ket{00} + \ket{11}}{\sqrt{2}})+\sin(\frac{\theta}{2})\qty(\frac{\ket{01} + \ket{10}}{\sqrt{2}})
\end{align}
with $\theta := -2\arccos\qty(1/\sqrt{1 + a^2})$ and $a := (U + \sqrt{U^2 + (8V)^2}) / (8V)$. This state can be efficiently created by the circuit 
\begin{equation}
    \text{CNOT}_{2,1}\cdot(\Ham_2 \otimes R_y(\theta)_1),
\end{equation}
where $\text{CNOT}_{1,2}$ denotes the controlled-NOT operator between qubits $1$ and $2$, as represented in the circuit below.

$$
\Qcircuit @C=1em @R=0.5em {
    \lstick{1}  &\gate{R_y(\theta)}           &\targ & \qw
    \\
    \lstick{2}  &\gate{H} &\ctrl{-1}    & \qw
    }
$$

\section{Simplifying Trotterization circuit \label{sec:trottercircuit}}

For the 1-site SIAM model, we use the 2$^{\rm nd}$ order Suzuki-Trotter formula, with two time steps. We implement the unitary
\begin{widetext}
\begin{equation}
    e^{-i\Ham t} \simeq \prod_{i=1}^2 e^{-i \frac{t}{4} \frac{V}{2} Z_a X_1} e^{-i \frac{t}{4} V X_2}  e^{-i \frac{t}{4} \frac{U}{4} Z_1 Z_2}  e^{-i \frac{t}{4} \frac{V}{2} X_1} e^{-i \frac{t}{4} \frac{V}{2} X_1} e^{-i \frac{t}{4} \frac{U}{4} Z_1 Z_2} e^{-i \frac{t}{4} V X_2} e^{-i \frac{t}{4} \frac{V}{2} Z_a X_1}. 
\end{equation}
\end{widetext}
A naïve implementation of this unitary and the rest of the circuit would require 19 CNOT gates (16 for the unitary, 1 for the ground state, and 2 to compute the Green's function). However, surprisingly, this circuit may be simplified down to only 2 qubits and 2 CNOT gates, making it amenable to current, and still very noisy, quantum hardware.

Collapsing the middle $VZ_aX_1$ exponentials reduces the CNOT count of the unitary from 16 down to 14. Furthermore, note that
\begin{widetext}
\begin{align}
   & e^{-i \frac{t}{4} \frac{U}{4} Z_1 Z_2}  e^{-i \frac{t}{2} \frac{V}{2} X_1} e^{-i \frac{t}{4} \frac{U}{4} Z_1 Z_2}\\
   ={}& \phantom{00} \Qcircuit @C=1em @R=0.5em {
    \lstick{1}  &\targ & \gate{R_z(-tU/8)} &\targ & \gate{R_x(-tV/2)} &\targ & \gate{R_z(-tU/8)} &\targ & \qw
    \\
    \lstick{2} &\ctrl{-1}    & \qw &\ctrl{-1}    & \qw &\ctrl{-1}    & \qw &\ctrl{-1}    & \qw
    }\\
    ={}& \phantom{00} \Qcircuit @C=1em @R=0.5em {
    \lstick{1}  &\targ & \gate{R_z(-tU/8)} & \gate{R_x(-tV/2)} & \gate{R_z(-tU/8)} &\targ & \qw
    \\
    \lstick{2} &\ctrl{-1}    & \qw  & \qw   & \qw &\ctrl{-1}    & \qw
    }
\end{align}
\end{widetext}
so the unitary CNOT count gets further reduced to 10. The full 13 CNOT gate circuit is then
\begin{widetext}
\begin{align}
    &\phantom{0}\scalebox{0.7}{
    \Qcircuit @C=0.2em @R=0.5em {
    \lstick{a} &\gate{H}    &\qw        &\ctrl{1}   &\qw        &\targ      &\gate{R_z} &\targ      &\qw        &\qw        &\qw        &\qw        &\qw        &\qw        
    &\qw        &\targ      &\gate{R_z} &\targ      &\qw        &\qw        &\qw        &\qw        &\qw        &\qw        
    &\qw        &\targ      &\gate{R_z} &\targ      &\qw        &\ctrl{1}   &\gate{H}   &\qw
    \\
    \lstick{1} &\gate{R_y}  &\targ      &\targ      &\gate{H}   &\ctrl{-1}  &\qw        &\ctrl{-1}  &\gate{H}   &\targ      &\gate{R_z} &\gate{R_x} &\gate{R_z} &\targ      
    &\gate{H}   &\ctrl{-1}  &\qw        &\ctrl{-1}  &\gate{H}   &\targ      &\gate{R_z} &\gate{R_x} &\gate{R_z} &\targ      
    &\gate{H}   &\ctrl{-1}  &\qw        &\ctrl{-1}  &\gate{H}   &\targ      &\qw        &\qw
    \\
    \lstick{2} &\gate{H}    &\ctrl{-1}  &\qw        &\qw        &\qw        &\gate{R_x} &\qw        &\qw        &\ctrl{-1}  &\qw        &\qw        &\qw        &\ctrl{-1}  
    &\qw        &\qw        &\gate{R_x} &\qw        &\qw        &\ctrl{-1}  &\qw        &\qw        &\qw        &\ctrl{-1}  
    &\qw        &\qw        &\gate{R_x} &\qw        &\qw        &\qw        &\qw        &\qw
    }
    }\\
    ={}&\phantom{0}\scalebox{0.7}{
    \Qcircuit @C=0.2em @R=0.5em {
    \lstick{a} &\gate{H}    &\ctrl{1}   &\qw        &\targ      &\gate{R_z} &\targ      &\qw        &\qw        &\qw        &\qw        &\qw        &\qw        &\qw        &\qw        
    &\qw        &\qw        &\qw        &\targ      &\gate{R_z} &\targ      &\qw        &\qw        &\qw        &\qw        &\qw        
    &\targ      &\gate{R_z} &\targ      &\qw        &\ctrl{1}   &\gate{H}   &\qw        &\qw
    \\
    \lstick{1} &\gate{R_y}  &\targ      &\gate{H}   &\ctrl{-1}  &\qw        &\ctrl{-1}  &\gate{H}   &\targ      &\qw        &\targ      &\gate{R_z} &\gate{R_x} &\gate{R_z} &\targ      &\qw        &\targ            
    &\gate{H}   &\ctrl{-1}  &\qw        &\ctrl{-1}  &\gate{H}   &\gate{R_z} &\gate{R_x} &\gate{R_z}      
    &\gate{H}   &\ctrl{-1}  &\qw        &\ctrl{-1}  &\gate{H}   &\targ      &\targ        &\qw      &\qw
    \\
    \lstick{2} &\gate{H}    &\qw        &\qw        &\qw        &\qw        &\qw        &\qw        &\ctrl{-1}  &\gate{R_x} &\ctrl{-1}  &\qw        &\qw        &\qw        &\ctrl{-1}  &\gate{R_x} &\ctrl{-1}
    &\qw        &\qw        &\qw        &\qw        &\qw  &\qw        &\qw        &\qw  
    &\qw        &\qw        &\qw        &\qw        &\qw        &\qw        &\ctrl{-1}  &\gate{R_x} &\qw
    }
    }
\end{align}
\end{widetext}
The final CNOT and $R_x$ are redundant, since we are only measuring the ancilla qubit. Therefore, the circuit is composed of consecutive 2-qubit blocks of 3, 4, and 5 CNOT gates. Each of these 2 qubit unitaries can be rewritten using just 2 CNOT gates, resulting in a circuit with 6 CNOT gates.

This circuit now has the form
\begin{widetext}
\begin{equation}
    \scalebox{1.0}{
    \Qcircuit @C=0.5em @R=0.5em {
    \lstick{a} &\gate{A_1}  &\targ      &\gate{A_3} &\targ      &\gate{A_5} &\qw        &\qw        &\qw        &\qw        &\qw        &\gate{C_1} &\targ      &\gate{C_3} &\targ      &\gate{C_5} &\qw   
    \\
    \lstick{1} &\gate{A_2}  &\ctrl{-1}  &\gate{A_4} &\ctrl{-1}  &\gate{A_6} &\gate{B_1} &\targ      &\gate{B_3} &\targ      &\gate{B_5} &\gate{C_2} &\ctrl{-1}  &\gate{C_4} &\ctrl{-1}  &\gate{C_6} &\qw
    \\
    \lstick{2} &\qw         &\qw        &\qw        &\qw        &\qw        &\gate{B_2} &\ctrl{-1}  &\gate{B_4} &\ctrl{-1}  &\gate{B_6} &\qw        &\qw        &\qw
    &\qw        &\qw        &\qw
    }
    }
\end{equation}
\end{widetext}
where the $A_i$, $B_i$, $C_i$ are single-qubit gates.
Again, note that the $B_6$ and $C_6$ gates do not need to be implemented. For our particular circuit, the gates $B_2$ and $B_4$ may be written solely using $R_z$ and $X$ gates. This means that qubit 2 is never in superposition (in the $Z$ basis) throughout the whole circuit. Therefore, the $R_z$ gates in this qubit are redundant, as they only add a global phase, and the $X$ gates may be pre-processed and removed. As a result, we know \emph{a priori} if the CNOT gates controlled by qubit 2 apply an $X$ gate or not to qubit 1, and we may replace them directly by these $X$ gates.

The final circuit, consequently, only has gates in the first two qubits. Rewriting this 2-qubit circuit once more, we end up with a circuit of the form
\begin{equation}\label{eq:circuit_run}
    \scalebox{1.0}{
    \Qcircuit @C=0.5em @R=0.5em {
    \lstick{a} &\gate{D_1}  &\targ      &\gate{D_3} &\targ      &\gate{D_5} &\qw
    \\
    \lstick{1} &\gate{D_2}  &\ctrl{-1}  &\gate{D_4} &\ctrl{-1}  &\gate{D_6} &\qw
    }
    }
\end{equation}
where the $D_6$ gate need not be applied. These gates will change as we change the value of $t$, but the circuit structure is always similar.

\section{Error mitigation \label{sect:error_mitigation}}

The measurements in the real quantum hardware are subject to various sources of noise, and so error mitigation techniques are desirable.
The goal is to obtain converted data that most closely matches the exact $\Gu(t)$.
For real data, the deviation from the correct values are mostly due to four error sources: the stochasticity of the measurements, time-evolution errors, readout errors, and gate errors.

We estimate expectation values by performing a finite number of measurements and taking the corresponding mean.
Therefore, our estimate has an associated variance.
Fortunately, the number of measurements necessary to reduce the error below $\epsilon_M$ is in general $\order{1/\epsilon_M^2}$, so it can be made arbitrarily small. 
Moreover, this error can be modeled as gaussian noise in the limit of many measurements, which is the setting assumed for the theoretical guarantees of the Atomic Norm Minimization \cite{MinMax, TaleOfResolution}.

We approximate the time evolution with a product formula.
The validity of this approximation worsens as we increase the evolution time.
So, the time-evolution routine contributes with a systematic bias to the computed $\Gu(t)$ if the expansion used does not have negligible error for the time domain considered.
We could increase the order of the product formula, but that would lead to a larger circuit and thus more decoherence errors.
In the end, we must bear in mind that the results are limited to the range where the product formula constitutes a good approximation of the time evolution operator.

The current devices do not measure the qubit registers with perfect fidelity.
To deal with these readout errors, we first compute a measurement filter, which indicates the observed deviation of the readout results from its expected values, for the particular hardware chip in use at the given time. 
With this filter, we can partially undo the effect of readout errors on our $\Gu(t)$ data. 
The measurement filter only needs to be computed once, and it can then be applied to all circuits run on that particular chip. 
As we are only interested in correcting the readout error of the ancilla qubit, this filter takes a constant, negligible time to compute, independent of the number of qubits or the circuits we intend to use.

For real quantum hardware, the gate's physical implementation is also not perfect. 
Its deviation from the desired gate introduces errors in the quantum state that may severely affect the output.
To mitigate these errors, we assume that the obtained noisy data $G_{\mu,\mathrm{noisy}}(t)$ is of the form
\begin{align}
    G_{\mu,\mathrm{noisy}}(t) &= \sum_l \tilde c_l e^{i2\pi f_l t},\quad \tilde c_l \simeq \alpha c_l,\quad \text{with }0 < \alpha \leq 1.
\end{align}
For $t=0$, we know from analytic considerations that $\Gu(0)=-i$.
By comparing the value of $\Gu(0)$ with the values of $\Gu(t)$ for small $t$, we estimate a value of $\alpha$.
Then, we rescale the measured data by $1 / \alpha$.
This admittedly \emph{ad-hoc} procedure effectively nullifies some effects of the encountered gate errors.
In general, we could also have implemented some form of zero noise extrapolation \cite{ZNE}, but we did not observe a significant advantage of doing so for our case.

\section{Device information \label{sect:hardware_configuration}}

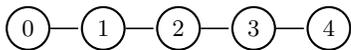
\begin{figure}[!htpb]
\centering
\begin{tikzpicture}[
        > = stealth, 
        shorten > = 1pt, 
        auto,
        node distance = 1cm, 
        semithick 
    ]

    \tikzstyle{every state}=[
        draw = black,
        thick,
        fill = white,
        minimum size = 4mm
    ]

    \node[state] (0) {$0$};
    \node[state] (1) [right of=0] {$1$};
    \node[state] (2) [right of=1] {$2$};
    \node[state] (3) [right of=2] {$3$};
    \node[state] (4) [right of=3] {$4$};

    \path[-] (0) edge node {} (1);
    \path[-] (1) edge node {} (2);
    \path[-] (2) edge node {} (3);
    \path[-] (3) edge node {} (4);
\end{tikzpicture}
\caption{Device layout. The nodes represent the qubits, and their labels, while the edges indicate the possible direct CNOT gate implementations.}
\label{fig:layout}
\end{figure}

\begin{table}[!htpb]
\begin{tabular}{l|cc}
\textbf{qubit}             & \textbf{3}   & \textbf{4} \\ \hline
$T_1$ ($\mu$s)             & 123.82       & 128.65     \\
$T_2$ ($\mu$s)             & 58.86        & 46.80      \\
frequency (GHz)            & 4.951        & 5.065      \\
anharmonicity (GHz)        & -0.344       & -0.342     \\
readout error (\%)         & 1.96         & 2.52       \\
$p(1|0)$ (\%)              & 0.74         & 1.30       \\
$p(0|1)$ (\%)              & 3.18         & 3.74       \\
readout length (ns)        & 5351         & 5351       \\
$\sqrt{X}$ infidelity (\%) & 0.045        & 0.045      \\
$\sqrt{X}$ duration (ns)   & 35.6         & 35.6       \\
CNOT(4,3) infidelity (\%)  & \multicolumn{2}{c}{0.68}  \\
CNOT(4,3) duration (ns)    & \multicolumn{2}{c}{298.7}
\end{tabular}
\caption{Device configuration.}
\label{tab:configuration}
\end{table}

The experimental results were obtained in IBM's \texttt{ibmq\_manila} device, in backend version 1.0.35. Qubits 3 and 4 were used to run the circuit in \Cref{eq:circuit_run} to obtain the results in \Cref{fig:1site_DFTANM}. Its layout and a representative configuration are displayed in \Cref{fig:layout,tab:configuration}, respectively. We note that, due to limitations in the number of circuits allowed to run concurrently in IBM's devices, and the large number of circuits required to obtain \Cref{fig:1site_DFTANM}, IBM's device may have been recalibrated between some of the runs. Nonetheless, the device configuration did not deviate significantly from the values shown in \Cref{tab:configuration}.

\end{document}